\documentclass[twocolumn]{el-author}
\usepackage{courier}
\usepackage{amsmath,amsthm,amsfonts,amssymb,bm}
\usepackage{algorithmic}
\usepackage{algorithm}
\usepackage{enumitem}
\usepackage{utfsym}
\usepackage{booktabs}
\usepackage{adjustbox}
\usepackage{threeparttable}
\usepackage{hyperref}
\usepackage{orcidlink}
\usepackage{graphicx} 
\usepackage{epstopdf}
\usepackage{marvosym}

\hypersetup{hypertex=true,
colorlinks=true,
linkcolor=blue,
anchorcolor=blue,
citecolor=blue}
\usepackage{booktabs} %
\usepackage{array} %

\newcommand{\ua}{\uparrow}
\newcommand{\nc}{\newcommand}
\nc{\da}{\downarrow} \nc{\hc}{\hat{c}} \nc{\hS}{\hat{S}}
\nc{\bra}{\langle} \nc{\ket}{\rangle} \nc{\eq}{equation (\ref}
\nc{\h}{\hat} \nc{\hT}{\h{T}}\nc{\be}{\begin{eqnarray}}
\nc{\ee}{\end{eqnarray}}\nc{\rd}{\textrm{d}}\nc{\e}{eqnarray}\nc{\hR}{\hat{R}}\nc{\Tr}{\mathrm{Tr}}
\nc{\tS}{\tilde{S}}\nc{\tr}{\mathrm{tr}}\nc{\8}{\infty}\nc{\lgs}{\bra\ua,\phi|}\nc{\rgs}{|\ua,\phi\ket}
\nc{\hU}{\hat{U}}\nc{\lfs}{\bra\phi|}\nc{\rfs}{|\phi\ket}\nc{\hZ}{\hat{Z}}\nc{\hd}{\hat{d}}\nc{\mD}{\mathcal{D}}
\nc{\bd}{\bar{d}}\nc{\bc}{\bar{c}}\nc{\mc}{\mathcal}\nc{\ea}{eqnarray}\nc{\mG}{\mathcal{G}}\nc{\bce}{\begin{center}}
\nc{\ece}{\end{center}}
\date{3rd October 2024}

\begin{document}

\title{HRRPGraphNet: Make HRRPs to Be Graphs for Efficient Target Recognition}

\author{Lingfeng Chen, Xiao Sun, Zhiliang Pan, Qi Liu, Zehao Wang, Xiaolong Su, Zhen Liu, Panhe Hu\textsuperscript{\Letter}}

\abstract{High Resolution Range Profiles (HRRPs) have become a key area of focus in the domain of Radar Automatic Target Recognition (RATR). Despite the success of deep learning based HRRP recognition, these methods needs a large amount of training samples to generate good performance, which could be a severe challenge under non-cooperative circumstances. Currently, deep learning based models treat HRRPs as sequences, which may lead to ignorance of the internal relationship of range cells. {This letter proposes HRRPGraphNet, a novel graph-theoretic approach, whose primary innovation is the use of the graph-theory of HRRP which models the spatial relationships among range cells through a range cell amplitude-based node vector and a range-relative adjacency matrix, enabling efficient extraction of both local and global features in noneuclidean space.} Experiments on the aircraft electromagnetic simulation dataset confirmed HRRPGraphNet's superior accuracy and robustness compared with existing methods, particularly in limited training sample condition. {This underscores the potential of graph-driven innovations in enhancing HRRP-based RATR, offering a significant advancement over sequence-based methods.} Codes are available at: \texttt{https://github.com/MountainChenCad/HRRPGraphNet}.}

\maketitle

\section{Introduction}
High Resolution Range Profiles (HRRPs), which are formed from the coherent summation of echoes emitted by target scattering centers, have emerged as a key asset in the domain of Radar Automatic Target Recognition (RATR). The physical characteristics of HRRPs, such as the structure and intensity of scattering centers, provide a rich vein of information for Radar Automatic Target Recognition (RATR)\cite{ref1, ref2}. Compared to Synthetic Aperture Radar (SAR) and Inverse SAR (ISAR), beyond its rich information of target structure and motion, HRRPs are easier to acquire and process, giving them a unique advantage in RATR\cite{ref*}.

Numerous HRRP RATR methods have been proposed in the past few decades. While Traditional HRRP recognition methods are hindered by their shallow architectural depth, limiting their performance\cite{ref3, ref9, ref10, ref**}, as a fast-growing technique, deep learning-based approaches methods achieve remarkable recognition results and have drawn great attention of the field\cite{ref4,ref5,ref6,ref7,ref8}. However, these models needs a considerable amount of data to achieve good performance, which could be a huge challenge especially under non-cooperative circumstances. HRRP RATR with limited training samples have therefore become a research hot-spot. 

Current HRRP RATR methods based on RNN, LSTM tend to focus on extracting global features while overlooking the structural characteristics of HRRP targets\cite{ref5, ref8}. Conversely, methods based on CNNs prioritize the extraction of local structural features but may not adequately consider global features or the relationships between these local structures\cite{ref4, ref12}. In general, current methods are sequenced-based, which might ignore the internal relationships between range cells that could hold the information of target structure, motion, etc. {Chen et al.\cite{ref13}, propose 1-DAMRAE which consider both local and global feature, yet its sequence-based LSTM context encoder requires a complex network structure.} 
\vspace{-1em}

\begin{figure}[h]
\centering
\includegraphics[width=3.45in]{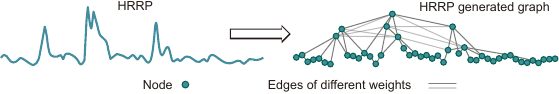}
\caption{A HRRP sequence and its graph representation}
\label{Fig_1}
\end{figure}

\vspace{-1em}
Inspired from structure-based Protein-Protein Interaction (PPI) prediction\cite{ref11}, this paper propose HRRPGraphNet, a novel graph neural network model that combines the advantages of noneuclidean graph-based data representation for HRRP recognition. {A key innovation, as Fig. \ref{Fig_1} illustrates, is the transformation of HRRP sequences into noneuclidean graphs, which adeptly leverages the spatial relationships between range cells for both local and global feature extraction and does not require a complex network structure compared with \cite{ref13}. Through specific design, local feature can be aggregated though a simple graph convolution layer, whose could possibly sets HRRPGraphNet apart from current sequence-based methods for better recogintion efficiency.}

\section{Proposed Method}\label{sec_2}The proposed HRRPGraphNet consists of three main parts: graph generation, the local and global feature extraction module, and the attention module. These modules work together to create an attentive graph convolution network specifically designed for our graph-theoretic approach for HRRP RATR. Initially, the input HRRP sequences are converted into graphs. Subsequently, the local and global features are extracted and fed into the attention module to produce the final result.

\vspace{-1em}
\subsection{Graph Generation}
Inspired from PPI prediction methods\cite{ref11}, which treat the whole protein sequence as a graph and classify the protein with graph classification methods. Graph model allow detailed consideration of local features and benefits the aggregation of global features. We designed a novel method to transform each HRRP sequence into a graph $\textit{\textbf{G}}=\{\textit{\textbf{V}}, \textit{\textbf{E}}\}$, which is consists of a node vector $\textit{\textbf{V}}$ and {an adjacency matrix} $\textit{\textbf{E}}$. For each input HRRP sample $\textit{\textbf{H}}=\{h_1, h_2, ..., h_N\}$, $N$ represents the dimension of the HRRP sequence and $h_1, h_2, ..., h_N$ represent the amplitude of $N$ range cells of the HRRP. Since each HRRP sequence has its corresponding graph, the range cells are considered as nodes of the graph, which can be expressed through the node vector $\textit{\textbf{V}}$ that can be calculated through $\textit{\textbf{V}}=\{h_{f}^{(1)}, h_{f}^{(2)}, ..., h_{f}^{(N)}\}$,
where, for the $i^{th}$ node on the graph, $h_{f}^{(i)}$ represents the multi-channels feature of embedded in the node. For the initially generated graphs, $\textit{\textbf{V}}=\textit{\textbf{H}}$ and $h_{f}^{(i)}=h_i$, which means each node is single channel with the amplitude of the range cell as pattern that serves as the input of the model.

In order to utilize the hidden information between the range cells, $\textit{\textbf{G}}$ is assumed as a fully-connected graph. Here, the relative range between the $i^{th}$ and the $j^{th}$ range cell is considered along with their amplitude, which are both vital information for target recognition. The weight of the edge $e_{i,j}$ between the $i^{th}$ and the $j^{th}$ can be defined as $e_{i,j}=\frac{h_{i}h_{j}}{|i-j|+1}$,
where, the product of cell amplitude $h_{i}$ and $h_{j}$ is divided by the sum of relative range $|i-j|$ and $1$ in case the denominator goes to $0$ when $i=j$. Subsequently, {the adjacency matrix} $\textit{\textbf{E}}$ can be defined through \eqref{equ_3}, a adjacent matrix formed by $e_{i,j}$:
\begin{equation}\label{equ_3}
\textit{\textbf{E}}=
\begin{bmatrix}
e_{1,1} &\cdots & e_{1,N} \\
\vdots & \ddots & \vdots \\
e_{N,1} & \cdots & e_{N,N}
\end{bmatrix}=
\frac{1}{|i-j|+1}\textit{\textbf{H}}^{T}\textit{\textbf{H}}
\end{equation}
For a HRRP sequence that has the size of $N$, the size of its adjacent matrix would be $N^2$. Here, a trick is that the calculation of $\textit{\textbf{E}}$ can be easily accomplished through low-rank vector $\textit{\textbf{H}}$ and its transpose $\textit{\textbf{H}}^{T}$. By this way, as shown in Fig. \ref{Fig_3}, it is evident that range cells in close proximity tend to have higher weights in the adjacency matrix $\textit{\textbf{E}}$ compared to those that are farther apart. Additionally, range cells with high amplitudes exhibit significantly higher edge weights with other cells, indicating their likelihood of high scattering center intensity. This underscores the physical relationships between range cells captured by our definition of $\textit{\textbf{E}}$ which could possibly achieve better feature representation and enhance the model recognition performance with limited training samples. 
\vspace{-1em}

\subsection{Local and Global Feature Extraction Module}
To aggregate both local and global feature of the HRRP sequence, two extraction modules for local and global feature extraction is specifically designed. The local feature extraction part is formed initially by two repeated 1-D convolution blocks, which is formed by a 1-D convolution layer with a kernel size of $1\times3$, followed by a batch normalization layer, the activation function is set to be LeakyReLU. For every single channel input $\textit{\textbf{V}}=\textit{\textbf{H}}$ that has the size of $1\times N$, the output channel size is set to be $D_{out}$, which means the output node vector $\textit{\textbf{V}}^{\prime}=\{h_{f}^{(1)}, h_{f}^{(2)}, ..., h_{f}^{(N)}\}$ has the size of $D_{out}\times N$. Till here, the graph generated by HRRP sequence is updated to a HRRP local feature-based graph $\textit{\textbf{G}}^{\prime}=\{\textit{\textbf{V}}^{\prime}, \textit{\textbf{E}}\}$. The extraction of local feature can be modeled as $\textit{\textbf{V}}^{\prime}=Conv_{1\times3}(Conv_{1\times3}(\textit{\textbf{V}}))$.
The global feature extraction module is formed by a graph convolution layer, which takes $\textit{\textbf{G}}^{\prime}$ as its input. Weight matrix $W_1\mathbb{\in}\mathbb{R}^{G_{out}\times D_{out}}$, $W_2\mathbb{\in}\mathbb{R}^{G_{out}\times D_{out}}$ and bias $b_i\mathbb{\in}\mathbb{R}^{G_{out}\times N}$ are learnable parameters of the layer. For the $i^{th}$ node $h_{f}^{(i)}$, the graph convolution can be expressed as \eqref{equ_5}:
\begin{equation}\label{equ_5}
h_{f,out}^{(i)}=W_{1}h_{f}^{(i)}+W_{2}\sum_{j\mathbb{\in} A(i)}e_{j, i}\cdot h_{f}^{(j)}+b_i
\end{equation}
where $j\mathbb{\in} A(i)$ represents the adjacent nodes of the $i^{th}$ node. The output of the layer is the node vector updated with global feature of the graph $\textit{\textbf{V}}_{out}=\{h_{out,f}^{(1)}, h_{out,f}^{(2)}, ..., h_{out,f}^{(N)}\}$, whose size is $G_{out}\times N$.

\vspace{-1em}
\subsection{Attention Module}
After the feature extraction, a attention module is designed to squeeze the feature channel and generate the classification result. Here, a linear attention model is introduced. The scoring function can be defined as \eqref{equ_7}:
\begin{equation}\label{equ_7}
s(h_{f,out}^{(i)})=h_{f,out}^{(i)~T}W_{att}+b_{att}
\end{equation}
where, weight matrix $W_{att}\mathbb{\in}\mathbb{R}^{G_{out}\times 1}$ and bias $b_{att}\mathbb{\in}\mathbb{R}^{G_{out}\times 1}$ are trainable parameters. The attention score is then distributed by Softmax function, which can be calculated through \eqref{equ_8}:
\begin{equation}\label{equ_8}
\textit{\textbf{V}}_{att}=\sum_{i=1}^{G_{out}}\alpha_{i}h_{f,out}^{(i)}=\sum_{i=1}^{G_{out}}\frac{h_{f,out}^{(i)}exp(s(h_{f,out}^{(i)}))}{\sum_{j=1}^{G_{out}}exp(s(h_{f,out}^{(j)}))}
\end{equation}
where, $\textit{\textbf{V}}_{att}\mathbb{\in}\mathbb{R}^{G_{out}\times 1}$ stands for the output of the attention layer. $\alpha_{i}$ stands for the attention distribution of the $i^{th}$ dimension. Then, $\textit{\textbf{V}}_{att}$ is processed through a fully connected layer, whose function can be formulated as $\textit{\textbf{V}}_{fc}=W_{fc}\textit{\textbf{V}}_{att}+b_{fc}$.
Here, weight matrix $W_{fc}\mathbb{\in}\mathbb{R}^{N\times C}$ and bias $b_{fc}\mathbb{\in}\mathbb{R}^{1\times C}$ are trainable parameters of the layer. $C$ stands for the number of targets to classify. The activation function is set to be log Softmax, after which the final result $c_{out}$ is obtained. The whole process can be modeled as $c_{out}=Fc(Att(\textit{\textbf{V}}_{out}))$.
The loss function $L$ of the model is cross entropy loss, which can be expressed as \eqref{equ_11}:
\begin{equation}\label{equ_11}
L = -\sum_{i=1}^{C}P^{*}_{i}\cdot log(P_{i})
\end{equation}
where, $P^{*}_{i}$ stands for the probability distribution of the $i^{th}$ class. $P_i$ represents the probability distribution generated for the $i^{th}$ class.

\vspace{-1em}
\section{Experiment}\label{sec_3}
To demonstrate the performance of HRRPGraphNet, experiments are conducted on the aircraft electromagnetic simulation dataset\cite{ref3}. In this section, the dataset is initially introduced. Subsequently, comparative experiments between HRRPGraphNet and the state-of-the-art HRRP RATR methods are performed. To explore the contribution of each module in HRRPGraphNet, ablation study is also carried out. To verify the efficacy of graph based methods compared to sequence-based methods, experiments under limited training sample scenario are considered. All the experiments are based on Linux through a desktop with Intel(R) Xeon(R) Bronze 3204 CPU and NVIDIA RTX A4000 GPU based on Pytorch 2.2.0 and Python 3.8.

\vspace{-1em}
\subsection{Experimental Datasets}
\begin{figure}[!t]
\centering
\includegraphics[width=3.35in]{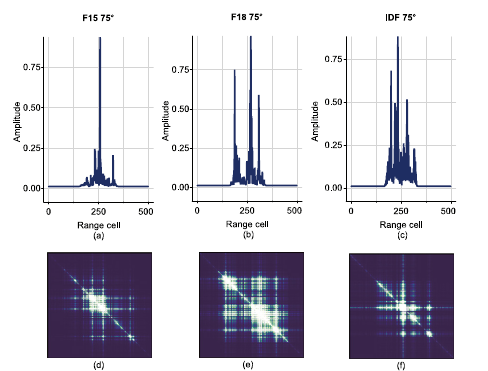}
\caption{The HRRP samples of different aircraft in of the $1^{st}$ pitch angel and the visualization of the corresponding adjacent matrix.}
\label{Fig_3}
\end{figure}

We simulated three classes of aircraft and obtained their radar echo signals with four polarization modes (HH, HV, VH, VV) based on the 3D aircraft model simulation software. The types of aircraft in the aircraft electromagnetic simulation dataset are F15, F18 and IDF (three different aircrafts), whose detailed parameters are shown in Table \ref{table_1}. The radar of the simulation system works in X band and the frequency range is $9.5$GHz\textasciitilde$10.5$GHz with step length of $10$MHz. The radar pitch angle is $75^{\circ}$\textasciitilde$105^{\circ}$ with step length of $3^{\circ}$ and the radar azimuth angle is $0^{\circ}$\textasciitilde$60^{\circ}$ with step length of $0.05^{\circ}$. Based on the above settings, to build a recognition scenario that is as realistic as possible, for each type of aircraft, we obtained a two sub-datasets with size of $2\times 900\times 501$ and $2\times 300\times 501$, $2$ means the number of pitch angles, $900$ and $300$ is the number of randomly chosen azimuth angles and $501$ means each HRRP sample contains $501$ range cells. The number $900$ and $300$ are set for experiments under few training sample scenario, whose training sample number is less than the number of the total azimuth angles $1201$. The training datasets $D_{train}$ is formed by the HH polarization mode $1^{st}$ pitch angle and all of the azimuth angles HRRPs, while the testing datasets $D_{test}$ is formed by the HH polarization mode $5^{th}$ pitch angle and the same number of the azimuth angles HRRPs, whose sizes are also $2\times 900\times 501$ and $2\times 300\times 501$. The HRRP sequences of the $1^{st}$ pitch angle is shown in Fig. \ref{Fig_3}.

\begin{table}[htbp]
  \centering
  \caption{The Parameters of Three Types of Aircraft in the Aircraft Elec-\\
  -tromagnetic Simulation Dataset.}
  \setlength{\tabcolsep}{4mm}{
    \begin{tabular}{cccc}
    \toprule
    Aircraft Type & Height($m$) & Length($m$) & Width($m$) \\
    \midrule
    F15   & $5.65$  & $19.45$ & $13.05$ \\
    F18   & $4.66$  & $17.07$ & $11.43$ \\
    IDF   & $4.70$  & $14.48$ & $8.53$ \\
    \bottomrule
    \end{tabular}}
  \label{table_1}%
\end{table}%

\begin{figure}[!t]
\centering
\includegraphics[width=2.8in]{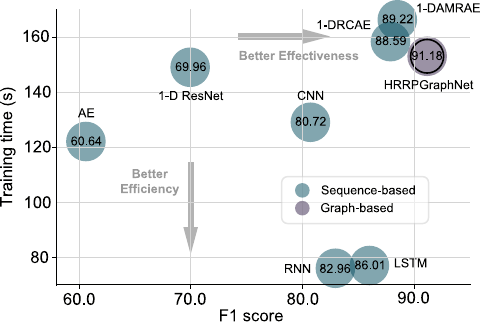}
\caption{F1 score \textit{vs.} Training time}
\label{Fig_4}
\end{figure}

\vspace{-1em}
\subsection{Comparative Experiments}
In this section, comparative experiments between proposed HRRPGraphNet and the existing methods are conducted on the the aircraft electromagnetic simulation dataset. {Comparative methods include traditional HRRP RATR methods (support vector machine (SVM)\cite{ref9}, linear discriminant analysis (LDA)\cite{ref10}, MSFKSPP-MMC\cite{ref**}), methods based on deep learning (recurrent neural network (RNN)\cite{ref8}, auto-encoder (AE)\cite{ref6}, long-short term memory (LSTM)\cite{ref5}, convolutional neural network (CNN)\cite{ref4}, 1-D ResNet\cite{ref7}, 1-DRCAE\cite{ref12}, 1-DAMRAE\cite{ref13}).} The comparative experiments results are shown in Table \ref{table_2}. {While traditional methods like LDA, and MSFKSPP-MMC generate competitive results, it can still be observed that our proposed method achieves the best recognition accuracy of $91.56\%$. when the number of training samples shrinks to $300$ per class.} Tables \ref{table_2}\textasciitilde \ref{table_3} shows that while other methods generally suffers from a decline of accuracy, our method maintains the best performance, achieving the highest recognition accuracy $90.78\%$. Considering efficiency and effectiveness as two important criteria, we report the F1 scores and the training times for 100 epochs of various deep learning RATR methods applied to the aircraft electromagnetic simulation dataset, as shown in Fig. \ref{Fig_4}. While simple methods exhibit efficiency in training time, they suffer from sub-optimal performance as they do not adequately model internal relationships between range cells. In contrast, our graph-based HRRPGraphNet can achieve fairly satisfactory performance by benefiting from the hidden information of HRRP structures, yet the computational burden is higher and could be improved in future research (e.g. a more efficient adjacency matrix). {Moreover, while 1-DAMRAE also takes both local and global feature, its computational burden is even higher due to its complicated network structure of its LSTM context encoder, which further proves the efficiency of HRRPGraphNet.}

\newpage
\begin{table}[htbp]
  \centering
  \caption{The Comparative Experimental Results on the Aircraft Electro-\\-magnetic Simulation Dataset with 900 Samples for Each Class.}
    \setlength{\tabcolsep}{3.1mm}{
    \begin{tabular}{ccccc}
    \toprule
    Method & F15   & F18   & IDF   & Average \\
    \midrule
    SVM\cite{ref9}   & $53.11$ & $42.22$ & $93.76$ & $63.17$ \\
    LDA\cite{ref10}   & $89.67$ & ${\textbf{99.89}}$ & $82.00$ & $90.52$ \\
    {MSFKSPP-MMC\cite{ref**}}   & ${89.00}$ & ${92.67}$ & ${86.33}$ & ${89.33}$ \\
    \midrule
    AE\cite{ref6}    & $80.33$ & $22.22$ & $91.56$ & $64.70$ \\
    CNN\cite{ref4}   & $91.67$ & $64.00$ & $92.22$ & $82.63$ \\
    RNN\cite{ref8}   & $59.00$ & $98.56$ & $96.44$ & $84.67$ \\
    1-D ResNet\cite{ref7} & $88.22$ & $28.78$ & ${\textbf{99.56}}$ & $72.19$ \\
    LSTM\cite{ref5}  & $65.44$ & $98.11$ & $95.67$ & $86.41$ \\
    {1-DRCAE\cite{ref12}}& ${72.67}$ & ${94.44}$ & ${98.89}$ & ${88.67}$ \\
    {1-DAMRAE\cite{ref13}}  & ${77.11}$ & ${95.11}$ & ${97.44}$ & ${89.89}$ \\
    \textbf{Ours} & ${\textbf{94.00}}$ & ${99.11}$ & $81.56$ & ${\textbf{91.56}}$ \\
    \bottomrule
    \end{tabular}}
  \label{table_2}%
\end{table}%
\vspace{-2em}
\begin{table}[htbp]
  \centering
  \caption{The Comparative Experimental Results on the Aircraft Electro-\\-magnetic Simulation Dataset with 300 Samples for Each Class.}
    \setlength{\tabcolsep}{3mm}{
    \begin{tabular}{ccccc}
    \toprule
    Method & F15   & F18   & IDF   & Average \\
    \midrule
    SVM\cite{ref9}   & $45.67$ & $39.00$ & $98.00$ & $60.89$ \\
    LDA\cite{ref10}   & $76.33$ & $98.67$ & $77.00$ & $84.00$ \\
    {MSFKSPP-MMC\cite{ref**}}   & ${89.33}$ & ${84.33}$ & ${81.67}$ & ${85.11}$ \\
    \midrule
    AE\cite{ref6}    & ${\textbf{93.67}}$ & $22.33$ & $89.67$ & $68.56$ \\
    CNN\cite{ref4}   & $48.67$ & $86.67$ & ${\textbf{98.67}}$ & $78.00$ \\
    RNN\cite{ref8}   & $53.33$ & ${\textbf{100.00}}$ & $93.00$ & $82.11$ \\
    1-D ResNet\cite{ref7} & $83.00$ & $16.00$ & $94.67$ & $64.56$ \\
    LSTM\cite{ref5}  & $78.00$ & $85.00$ & $92.67$ & $85.56$ \\
    {1-DRCAE\cite{ref12}} & ${90.67}$ & ${72.00}$ & ${93.33}$ & ${85.33}$ \\
    {1-DAMRAE\cite{ref13}}  & ${84.67}$ & ${81.67}$ & ${96.67}$ & ${87.67}$ \\
    \textbf{Ours} & $80.33$ & ${\textbf{100.00}}$ & $92.00$ & ${\textbf{90.78}}$ \\
    \bottomrule
    \end{tabular}}
  \label{table_3}%
\end{table}%
\vspace{-2em}
\begin{table}[htbp]
  \centering
  \begin{threeparttable}
  \caption{The Recognition Performance of the Proposed Modules in HR-\\-RPGraphNet ~on~ the Aircraft~ Electromagnetic Simulation Dataset with \\900 Samples for Each Class.}
    \setlength{\tabcolsep}{3mm}{
    \begin{tabular}{ccccccc}
    \toprule
    Number & a     & b     & c     & Accuracy & Recall & F1-score \\
    \midrule
    1     & \usym{2714} &       &       & $62.68$ & $62.55$ & $61.98$ \\
    2     &       & \usym{2714} &       & $83.22$ & $83.22$ & $82.22$ \\
    3     &       &       & \usym{2714} & $73.66$ & $73.51$ & $70.94$ \\
    4     & \usym{2714} & \usym{2714} &       & $84.09$ & $84.19$ & $83.59$ \\
    5     & \usym{2714} &       & \usym{2714} & $82.44$ & $82.51$ & $82.19$ \\
    6     &       & \usym{2714} & \usym{2714} & $65.37$ & $65.27$ & $65.28$ \\
    7     & \usym{2714} & \usym{2714} & \usym{2714} & ${\textbf{91.56}}$ & ${\textbf{91.31}}$ & ${\textbf{91.18}}$ \\
    \bottomrule
    \end{tabular}}
    \label{table_4}%
  \end{threeparttable}
\end{table}%
\vspace{-2em}


\subsection{Ablation Study}

As illustrated in Tables \ref{table_4}\textasciitilde \ref{table_5}, where a, b and c represent local feature extraction module, global feature extraction module and attention module respectively, by comparing experiment $2$ and $4$, $3$ and $5$, $6$ and $7$, it is clear that the proposed local feature extraction module improves the recognition accuracy. The accuracy with the local feature extraction module is $0.87\%$, $8.78\%$, $26.19\%$ higher than that without the local feature extraction module with $900$ samples for each class. When there are $300$ samples for each class, the gap was $1.89\%$, $3.67\%$, $4.78\%$. The results suggest that extracting local features before global feature extraction enhances the feature representation ability of the model, which helps the model to generalize when limited training samples are given.

Comparing experiment $1$ and $4$, $3$ and $6$, $3$ and $7$, it can be found that integrating either the local feature extraction module or the linear attention module with the global feature extraction module leads to a notable enhancement in the model's performance. Similarly, by conducting a comprehensive analysis of experiment $1$ and $5$, $2$ and $6$, $4$ and $7$, we found that the linear attention module boosts performance when combined with other parts of the network. It is also worth mentioning that comparing experiment 6 between Tables \ref{table_4}\textasciitilde \ref{table_5}, as the sample number in each class decreases, the contribution of the combined global feature extraction module and linear attention module appears to increase. In conclusion, our results proved that the every module proposed in this paper plays a crucial role in improving recognition performance and acts as a key component in the HRRPGraphNet.

\vskip-5pt
\section{Conclusion}\label{sec_4}
{This letter introduces HRRPGraphNet, a pioneering graph-theoretic approach for efficient HRRP RATR. The method overcomes the limitations of existing sequence-based deep learning approaches on effectiveness and efficiency by incorporating both local and global features of HRRP data into the recognition process through a novel transformation of HRRPs into graphs.} Extensive experiments conducted on aircraft electromagnetic simulation datasets have showcased the superior accuracy and robustness of HRRPGraphNet when compared to the exsiting methods, especially in scenarios with limited training samples. Limitations still exists, for instance, our current graph-based approach suffers from higher computational burden, which could still be a promising direction for future work.
\vspace{-1em}

\begin{table}[htbp]
  \centering
  \begin{threeparttable}
  \caption{The Recognition Performance of the Proposed Modules in HR-\\-RPGraphNet ~on~ the Aircraft~ Electromagnetic Simulation Dataset with \\300 Samples for Each Class.}
    \setlength{\tabcolsep}{3mm}{
    \begin{tabular}{ccccccc}
    \toprule
    Number & a     & b     & c     & Accuracy & Recall & F1-score \\
    \midrule
    1     & \usym{2714} &       &       & $64.22$ & $64.22$ & $64.69$ \\
    2     &       & \usym{2714} &       & $82.11$ & $82.11$ & $81.78$ \\
    3     &       &       & \usym{2714} & $68.89$ & $68.89$ & $63.52$ \\
    4     & \usym{2714} & \usym{2714} &       & $84.00$ & $84.00$ & $83.84$ \\
    5     & \usym{2714} &       & \usym{2714} & $72.56$ & $72.56$ & $70.21$ \\
    6     &       & \usym{2714} & \usym{2714} & $86.00$ & $86.00$ & $86.14$ \\
    7     & \usym{2714} & \usym{2714} & \usym{2714} & ${\textbf{90.78}}$ & ${\textbf{90.78}}$ & ${\textbf{90.70}}$ \\
    \bottomrule
    \end{tabular}}
    \label{table_5}%
  \end{threeparttable}       
\end{table}%

\vspace{-3em}
\ack{This work was supported by National Natural Science Foundation of China under Grant 62201588, in part by Natural Science Foundation of Hunan under Grant 2024JJ10007, 
in part by the Project funded by China Postdoctoral Science Foundation under Grant 2022M723914.}

\vskip5pt

\noindent Lingfeng Chen, Xiao Sun, Zhiliang Pan, Qi Liu, Zehao Wang, Xiaolong Su, Zhen Liu, Panhe Hu (\textit{College of Electronic Science and Technology, National University of Defense Technology, Changsha 410073, China})
\vskip3pt

\noindent \Letter~E-mail: hupanhe13@nudt.edu.cn

\vfill


\begin{thebibliography}{1}
\bibliographystyle{IEEEtran}

\bibitem{ref1}
{A. Aubry, V. Carotenuto, A. D. Maio, and L. Pallotta, “High range resolution profile estimation via a cognitive stepped frequency technique,” {\it{IEEE Trans. Aerosp. Electron. Syst.}}, vol. 55, no. 1, pp. 444-458, 2019.}

\bibitem{ref2}
{P. Addabbo, A. Aubry, A. D. Maio, L. Pallotta, and S. L. Ullo, “Hrr profile estimation using slim,” {\it{IET Radar, Sonar Navig.}}, vol. 13, no. 4, pp. 512-521, 2019.}

\bibitem{ref*}
{A. R. Persico, C. V. Ilioudis, C. Clemente, and J. J. Soraghan, “Novel classification algorithm for ballistic target based on hrrp frame,” {\it{IEEE Trans. Aerosp. Electron. Syst.}}, vol. 55, no. 6, pp. 3168-3189, 2019.}

\bibitem{ref**}
{X. Yang, G. Zhang, and H. Song, “Ship recognition based on hrrp via multi-scale sparse preserving method,” {\it{J. Syst. Eng. Electron.}}, vol. 35, no. 3, pp. 599-608, 2024.}

\bibitem{ref3}
Q. Liu, X. Zhang, and Y. Liu, “A prior-knowledge guided neural network based on supervised constrastive learning for radar hrrp recognition,” {\it{IEEE Trans. Aerosp. and Electron. Syst.}}, vol. 60, no. 3, pp. 1-21, 2024.

\bibitem{ref4}
J. Wan, B. Chen, B. Xu, H. Liu, and L. Jin, “Convolutional neural networks for radar hrrp target recognition and rejection,” {\it{EURASIP J. Adv. Signal Process}}, vol. 2019, no. 1, pp. 1–17, 2019.

\bibitem{ref5}
Q. Liu, X. Zhang, and Y. Liu, “Hierarchical sequential feature extraction network for radar target recognition based on hrrp,” in {\it{Proc. Int. Conf. Signal Image Process. (ICSIP)}}, 2022.

\bibitem{ref6}
Z. Fu, S. Li, B. Dan, and X. Wang, “A performance analysis of neural network models in hrrp target recognition,” {\it{Proc. IEEE Int. Conf. Signal Inf. Data Process. (ICSIDP)}}, 2019.

\bibitem{ref7}
J. Maitre, K. Bouchard, C. Bertuglia, and S. Gaboury, “Recognizing activities of daily living from uwb radars and deep learning,” {\it{Expert Syst. Appl.}}, vol. 164, p. 113994, 2021.

\bibitem{ref8}
Y. Zhang, F. Xiao, F. Qian, and X. Li, “Vgm-rnn: hrrp sequence extrapolation and recognition based on a novel optimized rnn,” {\it{IEEE Access}}, vol. 8, pp. 70071–70081, 2020.

\bibitem{ref9}
P. Guo, Z. Liu, and J. Wang, “Radar group target recognition based on HRRPs and weighted mean shift clustering,” {\it{ J. Syst. Eng. Electron.}}, vol. 31, no. 6, pp. 1152–1159, 2020.

\bibitem{ref10}
H. Yu and J. Yang, “A direct lda algorithm for high-dimensional data with application to face recognition,” {\it{Pattern Recognit.}}, vol. 34, no. 10, pp. 2067–2070, 2001.

\bibitem{ref11}
Z. Gao, C. Jiang, J. Zhang, X. Jiang, L. Li, P. Zhao, H. Yang, Y. Huang, and J. Li, “Hierarchical graph learning for protein–protein interaction,” {\it{Nat. Commun.}}, vol. 14, no. 1, pp. 1093, 2023b.

\bibitem{ref12}
{J. Yu and X. Zhou, “One-dimensional residual convolutional autoencoder based feature learning for gearbox fault diagnosis,” {\it{IEEE Trans. Industr. Inform.}}, vol. 16, no. 10, pp. 6347-6358, 2020.}

\bibitem{ref13}
{K. Chen, J. Zhang, S. Chen, S. Zhang, and H. Zhao, “Active jamming mitigation for short-range detection system,” {\it{IEEE Trans. Veh. Technol.}}, vol. 72, no. 9, pp. 11446-11457, 2023.}

\end{thebibliography}
\end{document}